\input{aipcheck}

\documentclass[
    ,final            % use final for the camera ready runs
  ]
  {aipproc}

\layoutstyle{6x9}

\def\slashed{\ds}

\usepackage{slashed}
\usepackage{graphics}
\usepackage{epsf}

\newcommand{\bea}{\begin{eqnarray}}
\newcommand{\eea}{\end{eqnarray}}
\newcommand{\ba}{\begin{eqnarray}}
\newcommand{\ea}{\end{eqnarray}}
\newcommand{\bann}{\begin{eqnarray*}}
\newcommand{\eann}{\end{eqnarray*}}
\newcommand{\bmi}{\begin{minipage}}
\newcommand{\emi}{\end{minipage}}

\newcommand{\beqa}{\begin{eqnarray}}
\newcommand{\eeqa}{\end{eqnarray}}
\def\beq{\begin{equation}}
\def\eeq{\end{equation}}

\begin{document}

\title{Gauged Axions and their QCD Interactions}


\author{Claudio Corian\`o}{
  address={Dipartimento di Fisica, Universit\`a del Salento 
and INFN Sezione di Lecce, Via Arnesano 73100 Lecce, Italy}
}

\author{Marco Guzzi}{
  address={Department of Physics, Southern Methodist University, Dallas, TX 75275-0175, U.S.A.}
}

\author{Antonio Mariano}{
  address={Dipartimento di Fisica, Universit\`a del Salento 
and INFN Sezione di Lecce, Via Arnesano 73100 Lecce, Italy},
%altaddress={} % additional visiting address
}

\begin{abstract}
 We present a brief overview of axion models associated to anomalous 
abelian (gauge) symmetries, discussing their main phenomenological features. 
Among these, the mechanism of vacuum misalignment introduced at the QCD and 
at the electroweak phase transitions, with the appearance of periodic potentials, 
responsible for the generation of a mass for these types of axions. 

\end{abstract}

\maketitle

%%%%%%%%%%%%%%%%%%%%%%%%%%%%%%%%%%%%%%%%%%%%
%% MAINMATTER
%%%%%%%%%%%%%%%%%%%%%%%%%%%%%%%%%%%%%%%%%%%%

\section{Introduction}
Axions were introduced in 1977 by Peccei and Quinn (PQ) as an attempt to solve the strong CP problem, which finds its origin in the smallness of the QCD $\theta$ angle, as the measurements of the electric dipole moment of the neutron confirm. The invisible axion, which emerged from a modification of the original PQ proposal, is expected to play a considerable role in cosmology, since it could be a significant component of cold dark matter (see \cite{Sikivie:2006ni}). For this reason they have been at the center of a large number of investigations,  in an attempt to identify possible ways of experimental detection of these particles.  

The generation of a PQ axion follows a well-established theoretical pattern, characterized  by the breaking of a global 
$U(1)_{PQ}$ symmetry at a very large scale $f_a$ (about $10^{12}$ GeV), followed by a tilting of the PQ potential, associated to this symmetry, and taking place at the quark-hadron transition. The mass of the axion ($m_a$) is indeed defined in terms of the two scales which characterize the first breaking ($f_a$) and the tilting $(\Lambda_{QCD})$ as 
$m_a\sim \Lambda_{QCD}^4/f_a^2$. The ``extra potential'' which causes the tilting of the PQ mexican-hat is attributed to 
QCD instantons. It is periodic, with a period given by $\pi f_a$ and a strength which is of the order of $\Lambda_{QCD}$. 
We just recall that an axion undergoes a global shift $(\theta\rightarrow \theta + \alpha)$ under a global $U(1)_{PQ}$ transformation, with $\alpha$ denoting the gauge parameter. For a gauged axion $\alpha$ is promoted to a 
local parameter. 

There have been several attempts to modify the standard PQ picture, mostly for cosmological reasons, for instance to provide a possible explanation of the dark energy problem, by introducing a global $U(1)$ symmetry 
broken at a very large scale as in PQ, but with a tilting occurring at the electroweak scale (see for instance  \cite{Nomura:2000yk}). In this case one needs to construct an anomalous (global) $U(1)$ symmetry characterized by a mixed anomaly just with the electroweak sector. Also in this 
case the axion, which remains a Nambu-Goldstone mode until a tilting in the broken original potential takes place, acquires a small mass. In the electroweak case, however, this turns out to be much smaller than the one obtained in the standard QCD scenario, due to the far smaller size of the ``extra potential'', which is due to electroweak instantons, compared to the QCD case.  

As we extend the theory and require the gauging of the anomalous $U(1)$ symmetry, the gauge invariance of the effective action requires the presence of local counterterms of Wess-Zumino type to restore the gauge symmetry, broken at 1-loop level. The cancellation of the anomaly, if realized by a local counterterm, indeed defines an anomaly-free theory with the appearance of an axion in the spectrum \cite{Coriano':2005js}. This new degree of freedom is described by a real St\"uckelberg field $b$ which undergoes a local shift under the anomalous gauge symmetry.

Theories of this type have clearly a unitarity bound \cite{Coriano:2008pg} but are consistent effective theories up to a certain scale which is given by the St\"uckelberg mass ($M_{St}$). This appears in two ways: 1) as a tree-level contribution to the mass of the anomalous gauge boson (via the St\"uckelberg mechanism) and as a suppression factor of the Wess-Zumino interactions 
($\sim b/M_{St} F\tilde F$), where $b$ is the St\"uckelberg field. Details on the construction of complete extensions of the 
Standard Model with one extra $U(1)$ gauge symmetry have been given in \cite{Coriano:2007xg}, to which we refer for further details. 

One additional remark concerns the possible motivations in favour of these types of theories, which are not justified only within models of intersecting branes, which originally brought to define their structure \cite{Coriano':2005js}. Their formulation is indeed rather general and therefore they grasp the essential features of theories with anomalous gauge interactions. Thus, they can be formulated in a bottom-up approach quite straightforwardly, just using some few basic and simple requirements, first among them being, indeed,  gauge invariance.  Notice that the St\"uckelberg field is dynamical, and this forces necessarily the anomalous gauge boson to acquire a mass at tree-level via a St\"uckelberg mass term. 

The discussion of simple models containing these essential features is contained in  \cite{Coriano:2007fw}. In principle, one could extend this construction to the supersymmetric case and also merge them into a standard gravitational framework. Indeed one may expect that these descriptions, after suitable generalizations, should grasp, at least in part, the physical features of theories with gauged supergravities 
(see for instance \cite{DeRydt:2007vg, Derendinger:2007xp}) which also contain axionic symmetries.  

As we have mentioned, one of the open issues, from the phenomenological viewpoint, concerns the strength of the extra potential, whose structure can be predicted on the basis of gauge invariance up to an overall factor which is left, however, undetermined. In this respect one can envision various possible ranges for the size of the mass of the physical axion that emerges from the St\"uckelberg field. The physical (gauged) axion, called the ``axi-Higgs'' in \cite{Coriano':2005js}, appears 
in the CP-odd sector after electroweak symmetry breaking and acquires a mass which depends on the size of the extra potential. 

The first window for the mass parameter is in the few GeV range, where the axion is akin a light pseudoscalar Higgs particle, with a fast decay and a phenomenology which has been studied in \cite{Coriano:2009zh}. A second window could correspond to a particle whose mass is in the keV/MeV region, heavy enough not to be produced at the center of the sun, thereby satisfying the current bounds from experiments such as CAST, which look directly for axions coming from the sun. 
A third possibility is a particle of a mass in the milli-eV region, which coincides with the expected mass range for a traditional 
PQ axion. In this case, the analysis of the cosmological implications of this particle and of the sequential mechanisms of misalignments which characterize its mass have been studied in \cite{Lazarides:1982tw}. 

\subsection{The phase-dependent potential}
One of the most significant features of axion models is in the appearance of a phase-dependent (or axion-dependent) potential which is periodic in the axion field and which can be inferred just on the basis of symmetry \cite{Coriano':2005js}. The simplest realization of models characterized by this feature involves gauge theories containing two Higgs doublets, that we are going to summarize. 
The model originally studied in \cite{Coriano':2005js, Coriano:2007xg} contains a single extra anomalous $U(1)$ plus an axion $b$ (a real pseudoscalar) to enforce cancellation of the anomaly.  

As in previous works \cite{Coriano:2007xg}, the scalar sector of the anomalous abelian models that we are interested in is characterized by an electroweak potential  involving, in the simplest formulation, two Higgs doublets $V_{P Q}(H_u, H_d)$ plus one extra contribution (the extra potential mentioned above), denoted as $V_{\slashed{P}\slashed{Q}}(H_u,H_d,b)$ or $V^\prime$, \cite{Coriano':2005js}  which mixes the Higgs sector with the St\"uckelberg axion $b$, needed for the restoration of the gauge invariance of the effective Lagrangian
\begin{equation}
V=V_{PQ}(H_u,H_d) + V_{\slashed{P}\slashed{Q}}(H_u,H_d,b).
\end{equation}
The appearance of the physical axion in the spectrum of the model takes place after 
that the phase-dependent terms, here assumed to be of non-perturbative origin and generated at the electroweak phase transition, find their way in the dynamics of the model and induce a curvature on the scalar potential. To better illustrate this point, we begin our analysis by turning to the ordinary potential of 2 Higgs doublets,
\ba
V_{PQ}&=&\mu_u^2 H_u^{\dagger}H_u+\mu_d^2 H_d^{\dagger}H_d+\lambda_{uu}(H_u^{\dagger}H_u)^2
+\lambda_{dd}(H_d^{\dagger}H_d)^2
\nonumber\\
&-&2 \lambda_{ud}(H_u^{\dagger}H_u)(H_d^{\dagger}H_d)
+2\lambda^{\prime}_{ud}\vert H_u^T \tau_2 H_d\vert^2 
\eeqa
to which we add a second term
\ba
V_{\slashed{P}\slashed{Q}}&=&\lambda_0(H_u^{\dagger}H_d e^{-i g_B (q_u-q_d)\frac{b}{2 M}})+
\lambda_1(H_u^{\dagger}H_d e^{-i g_B (q_u-q_d)\frac{b}{2 M}})^2
\nonumber\\
&+&\lambda_2(H_u^{\dagger}H_u)(H_u^{\dagger}H_d e^{-i g_B (q_u-q_d)\frac{b}{2M}})+
\nonumber\\
&&\lambda_3(H_d^{\dagger}H_d)(H_u^{\dagger}H_d e^{-i g_B (q_u-q_d)\frac{b}{2 M}})+\textrm{h.c.}.
\ea
These terms, as we have mentioned, are allowed by the symmetry of the 
model and are parameterized by one dimensionful  ($\lambda_0$) and three dimensionless constants ($\lambda_1,\lambda_2,\lambda_3$). They are assumed to be generated at the electroweak phase transition. The physical axion $\chi$ emerges as a linear combination of the phases of the various terms, which are either due to the CP-odd components of the Higgs sector or to the St\"uckelberg field $b$. A careful analysis of the the CP-odd sector shows that 
we can construct the orthogonal matrix $O^\chi$ that rotates the 
St\"uckelberg field and the CP-odd phases of the two Higgs doublets into the mass eigenstates 
$(\chi, G^{\,0}_1, G^{\,0}_2)$
\bea
\left(
\begin{array}{c} ImH^{0}_{u} \\
                 ImH^{0}_{d} \\
					  b
\end{array} \right)=O^{\chi}
\left(
\begin{array}{c} \chi\\
                 G^{0}_1\\
					  G^{0}_2
\end{array} \right)
\label{CPodd}
\eea
of the same sector.
The mass matrix of this sector exhibits two zero eigenvalues 
corresponding to the Goldstone modes $G^{0}_1, G^{0}_2$ and a nonzero eigenvalue, 
that corresponds to the physical axion field $ \chi$, with a mass given by
\beq
m_{\chi}^2=\frac{2 v_u v_d}{\sigma^2_\chi}\left(\bar{\lambda}_0 v^2 +\lambda_2 v_d^2 +\lambda_3 v_u^2+4 \lambda_1 v_u v_d\right) 
\approx \lambda v^2
\label{axionmass}
\eeq
( with $\lambda_0=\bar{\lambda}_0 v^2$).
Notice that one can rewrite the extra potential in the form\cite{Coriano:2010py}

\beq
V^\prime= 4 v_u v_d
\left(\lambda_2 v_d^2+\lambda_3 v_u^2+\lambda_0\right) \cos\left(\frac{\chi}{\sigma_\chi}\right) + 2 \lambda_1 v_u^2 v_d^2 \cos\left(2\frac{\chi}{\sigma_\chi}\right),
\label{extrap}
\eeq

with $\chi$ a gauge invariant degree of freedom. As we have remarked, this state is extracted from the CP-odd sector after a careful study of the Goldstone modes of the theory in the broken electroweak phase. In (\ref{extrap})  $\sigma_\chi$ is of the order of the electroweak scale and the ratio $\chi/\sigma_\chi$ is clearly an angle of misalignment.
From a simple phenomenological point of view, given the presence of a free parameter in the characterization of the extra potential, one can envision several possible scenarios. In \cite{Coriano:2010py} it has been assumed that this potential can be generated non perturbatively by some instanton interaction, and as such its strength $(\sim \lambda)$ should be strongly suppressed, but obviously there are other options. The one which is more interesting for LHC studies involves a light axion which is Higgs-like, but with a mass that covers the few GeV region. In the next section we will briefly explore this possibility, showing some predictions for the main channels of production and decay of this particle at the new collider.

\section{Axi-Higgs decay and production at hadron colliders}

\begin{figure}[h]
%\begin{center}
%\subfigure[]
{\includegraphics[
width=6cm,
angle=-90]{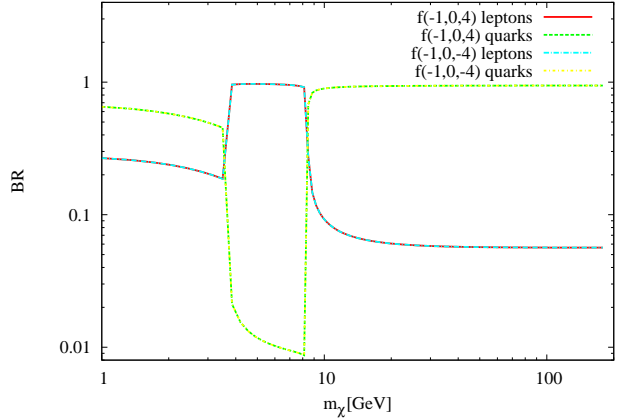}}
\caption{\small Study of the leptonic and the quarks branching ratios of the axi-Higgs. 
We analyze the dependence on the function $f(q^B_{Q_L},q^B_{u_R},\Delta q^B)$.} 
\label{Madrid_model}
%\end{center}
\end{figure}

We show in Fig.~(\ref{Madrid_model}) plots which illustrate the behaviour of the 
(inclusive) decay branching ratios of the axi-Higgs into quarks and leptons as a function of the mass of the physical axion, obtained by varying the charge assignments of the model.  The enhancement of the lepton decay channels for a light axion mass between 4 and 8 GeV - respect to the quark channel - is quite evident from this plot, and the results are rather  
stable respect to the variations induced by the different charge assignments of the model. These have been parameterized by the function $f(q^B_{Q_L},q^B_{u_R},\Delta q^B)$ of three variables. These are the charges of the left and right-handed quarks under 
the anomalous $U(1)_B$ gauge group ($q^B_{Q_L},q^B_{u_R}$)  while $\Delta q^B$ denotes the difference in the charges of the two Higgses under the same gauge group. The remaining charges of the model are fixed by anomaly cancellation and are described in \cite{Coriano:2009zh}. Notice that the differences among the various charge assignments are 
smaller than $10^{-3}$.

The study of the production of the axi-Higgs at hadron colliders is particularly interesting, especially for the possibility of having sizeable branching ratios of the two CP-even Higgs Higgs $H_0$ and $h_0$ into final state axions.  
 
The free parameters of the scalar potential can be identified by the coefficients
$(\lambda_{uu},\lambda_{dd},\lambda_{ud},\lambda'_{ud})$ that are contained in the
$PQ$ potential and by $(b_1,\lambda_1,\lambda_2,\lambda_3)$, that are contained in
the $PQ$-breaking potential.
The other free parameters are the ratio of the Higgs vevs, identified with $\tan\beta$,
the St\"uckelberg mass $M_{St}$ and the coupling constant $g_B$.

We start our analysis by considering a scenario in which the mass
of the $Z$ boson is exactly reproduced at $M_Z=91.1876$ GeV and the
bounds on the mass of the extra $Z'$ are required to be compatible with the current
Tevatron data. These conditions can be obtained by
fixing the value of the anomalous coupling $g_B\approx 0.1$, the value of $v_u\approx 246$ GeV,
the value of the Stueckelberg mass in the TeV range and $\tan\beta=40$.
These requirements induce also a small mixing parameter between $Z$ and $Z'$ (below $10^{-3}$), which is also in agreement with current data.
Thus, the mass of the particles of the scalar sector are identified
by the eight parameters listed above.
The value of the mass of the axi-Higgs is completely governed by the $PQ$-breaking
sector of the potential.
The other parameters enter in the definition of the mass of the two neutral Higgs
and the corresponding mass eigenvalues are found to be very sensitive to the selection of these parameters. 
In our case these have been chosen as $\left\{\lambda_1,\lambda_2,\lambda_3,b_1,\lambda_{uu},\lambda_{dd},\lambda_{ud}\right\} =\left\{-9~10^{-5},-1~10^{-6},-1~10^{-5},5~10^{-3},6~10^{-2},5,0.9\right\}$, with the  masses of the CP-even and of the CP-odd
sectors given by $\left\{m_{H_0}\approx 122,m_{H_0}\approx 15,m_{\chi}\approx 5\right\}$ (GeV).

\begin{figure}[t]
%\begin{center}
\includegraphics[width=6cm, angle=0]{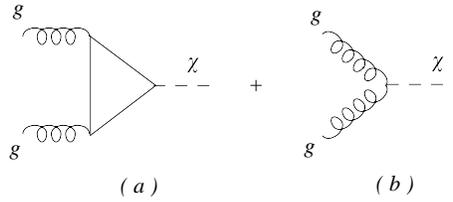}
\caption{\small The two contributions to the $gg \rightarrow \chi$ production channel.
\label{chi_production}}
%\end{center}
\end{figure}

\begin{figure}[t]
\includegraphics[width=5cm, angle=-90]{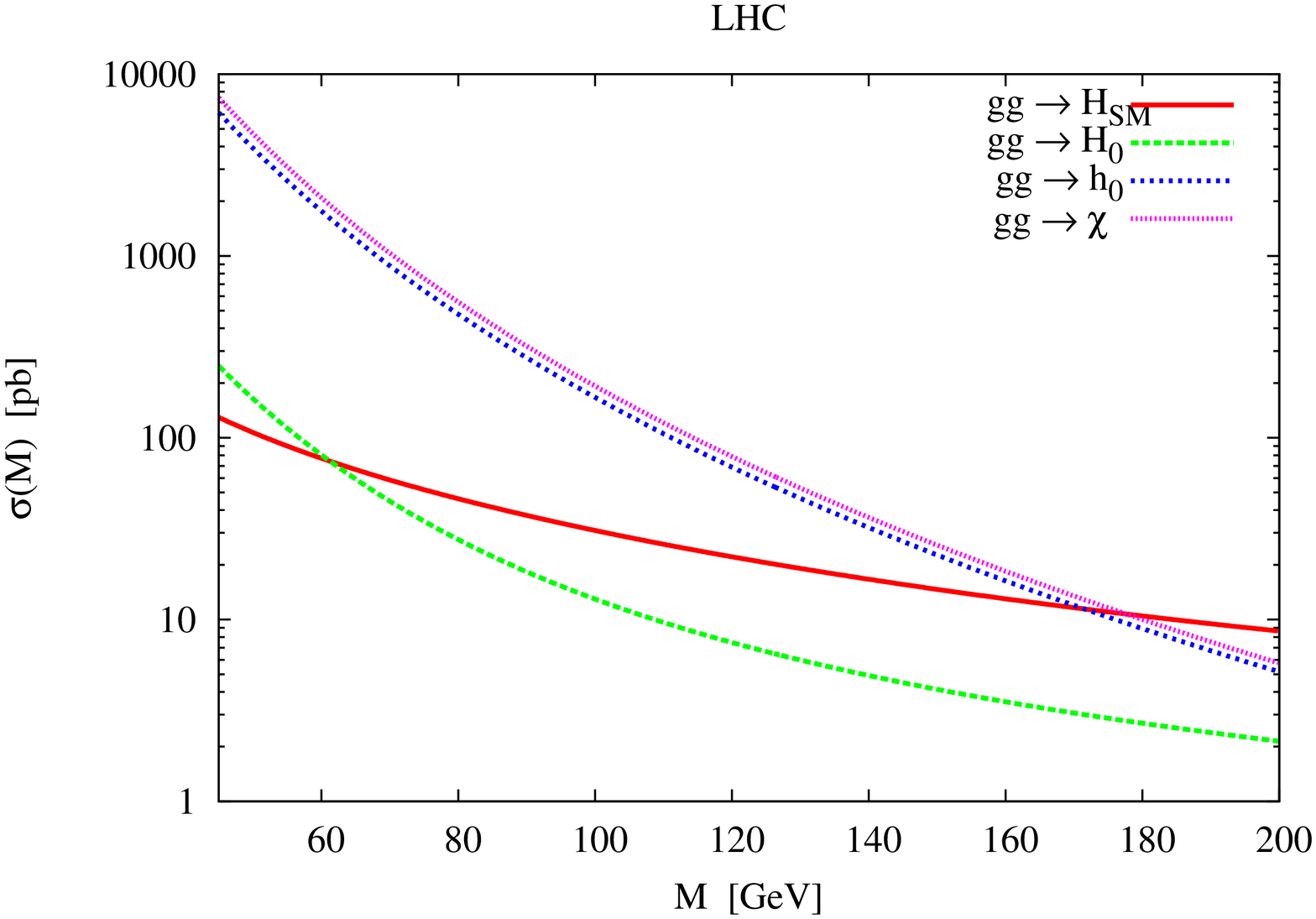}
\includegraphics[width=5cm, angle=-90]{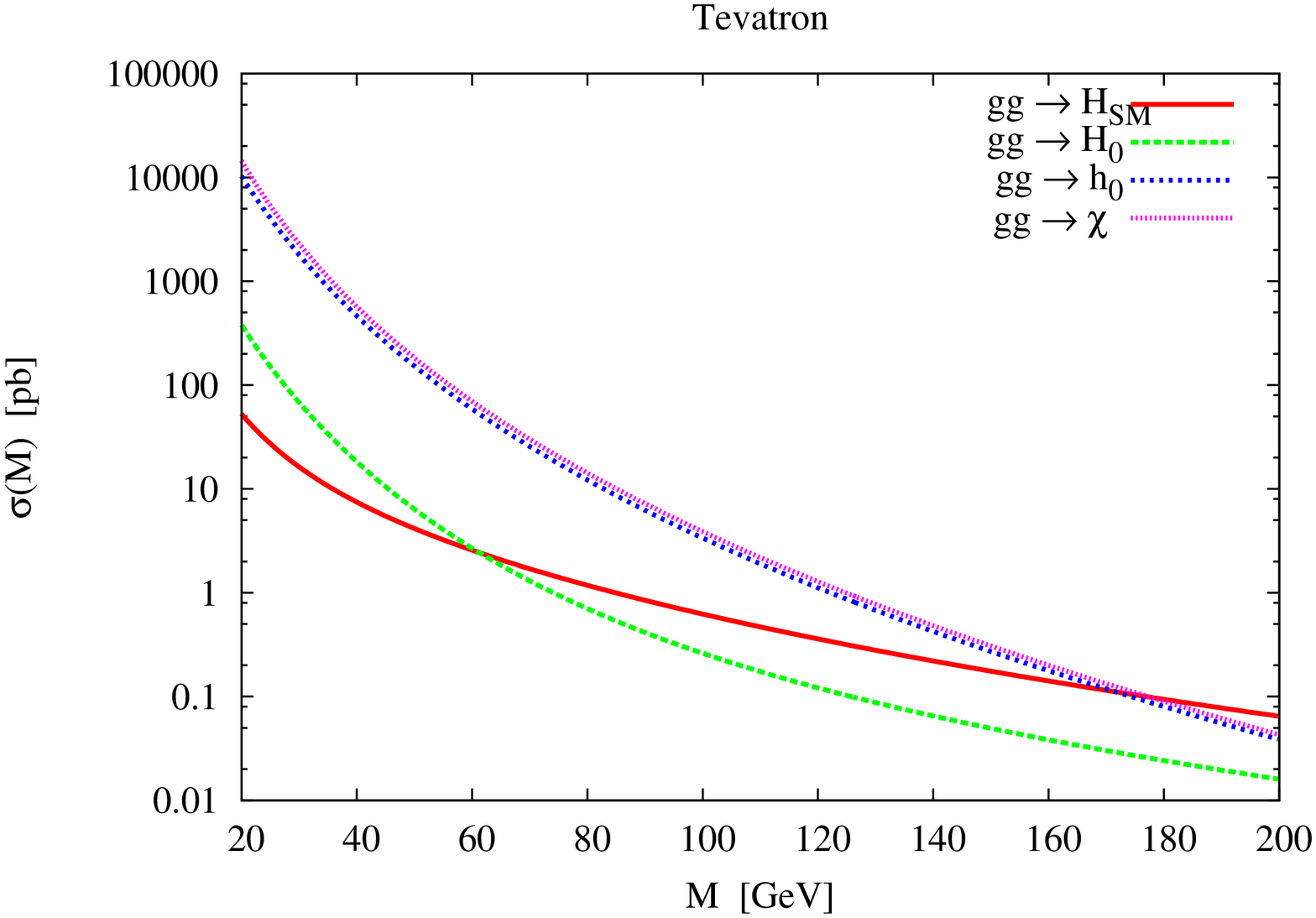}
\caption{\small  Cross section for the production of the two Higgs 
$h_0$ and $H_0$ and the axi-Higgs via gluon-gluon fusion at LO at the 
LHC (left panel) and at the Tevatron (right panel).}
\label{compares}
\end{figure}

In Figs. \ref{compares} we show the plots of the total
cross section at leading order (LO) - at the LHC and at the Tevatron respectively - for the production
of the axion and the CP-even $H_0,h_0$ Higgs, and the corresponding plots for the SM Higgs. Notice that the result shows a rather sharp rise of the production cross section as the axion mass gets lighter. This is larger by a factor of 10 compared to the case of other CP-even scalars. A similar rise is found also for the CP-odd sector in 2 Higgs doublet models, being a feature typical of the CP-odd Higgs sector.

\subsection{Axion plus photon production}

\begin{figure}[t]
%\begin{center}
\includegraphics[width=9cm]{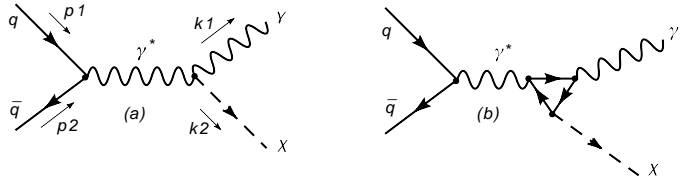}
\caption{\small Associated production channel for a single gauged axion plus a photon}
\label{X_prod}
%\end{center}
\end{figure}

\begin{figure}[t]
%\begin{center}
\includegraphics[width=6cm,angle=-90]{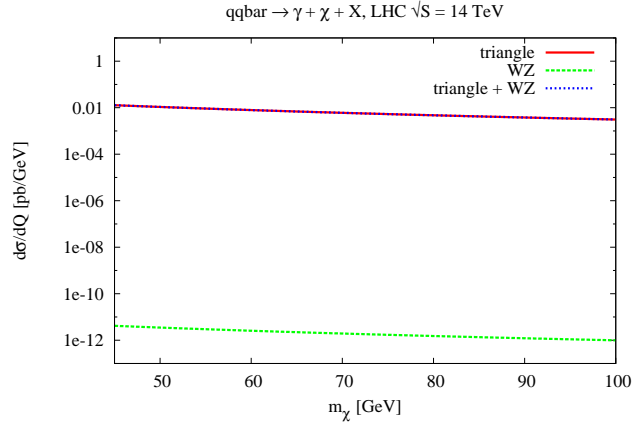}
\caption{\small Invariant mass distribution for the associated production of an
axion plus one photon at the LHC.
\label{crossXph}}
%\end{center}
\end{figure}

The diagrams responsible for associated production of a gauged axion and a photon are shown in Fig. \ref{X_prod}. We show instead in Fig.~\ref{crossXph} a plot of the cross section for the production of an axion and one photon at the LHC as a function of the mass of $\chi$. The mass dependence of the result is quite small, except at 
larger mass value of $\chi$, in a region where it is Higgs-like. For an ultralight axion, instead, the value of the cross section is around $10^{-2}$ pb. We have shown the contribution from the triangle and the Wess-Zumino terms combined together and separately, in order to show the dominance of one channel respect to the other. The Wess-Zumino term is indeed strongly suppressed (by a factor of $10^{10}$) respect to the fermion loop. 

\subsection{Multi axion production}
One of the peculiarities of a light axion-like particle is its possibility to generate cascade decays with multi-lepton final states which become more sizeable especially for a mass of $\chi$ in the GeV range. For 
$m_\chi$ around few GeV's the largest contribution to the decay branching ratio comes from the lepton channel, and for this reason we have investigated this particular interval in parameter space. 

The sizeable values of the invariant mass distributions for multi-axions in the final state are related to the large production cross sections which are typical of pseudoscalar channels, one of the reasons being the large values
of the reduced couplings - normalized to the SM ones - of the trilinear interactions
of the scalars. The leading contribution to the production cross section comes from the fermion loop graph 
with a final state axion. In the model each contribution is accompanied by the corresponding WZ counterterm, which is suppressed by a factor of $10^{5}$ compared to the loop graph (see Fig. \ref{chi_production}). 

Channels involving several final state axions can be built rather easily. For instance, the simplest process involves a 
$gg-h_0$ (two gluon $g$) production channel combined with the $h_0-\chi\chi$ vertex. In this case the WZ counterterm is absent. 
A similar process is the $gg-\chi$ triangle vertex, followed by the $\chi\chi h_0$ vertex, which gives the combination of a $\chi$ and of a CP-even Higgs $(h_0)$ in the final state. In this case the channel is accompanied by a WZ term $gg-\chi$ describing the direct interaction of the two gluons of the initial state with the axion. 

\begin{figure}[t]
\includegraphics[width=6cm, angle=-90]{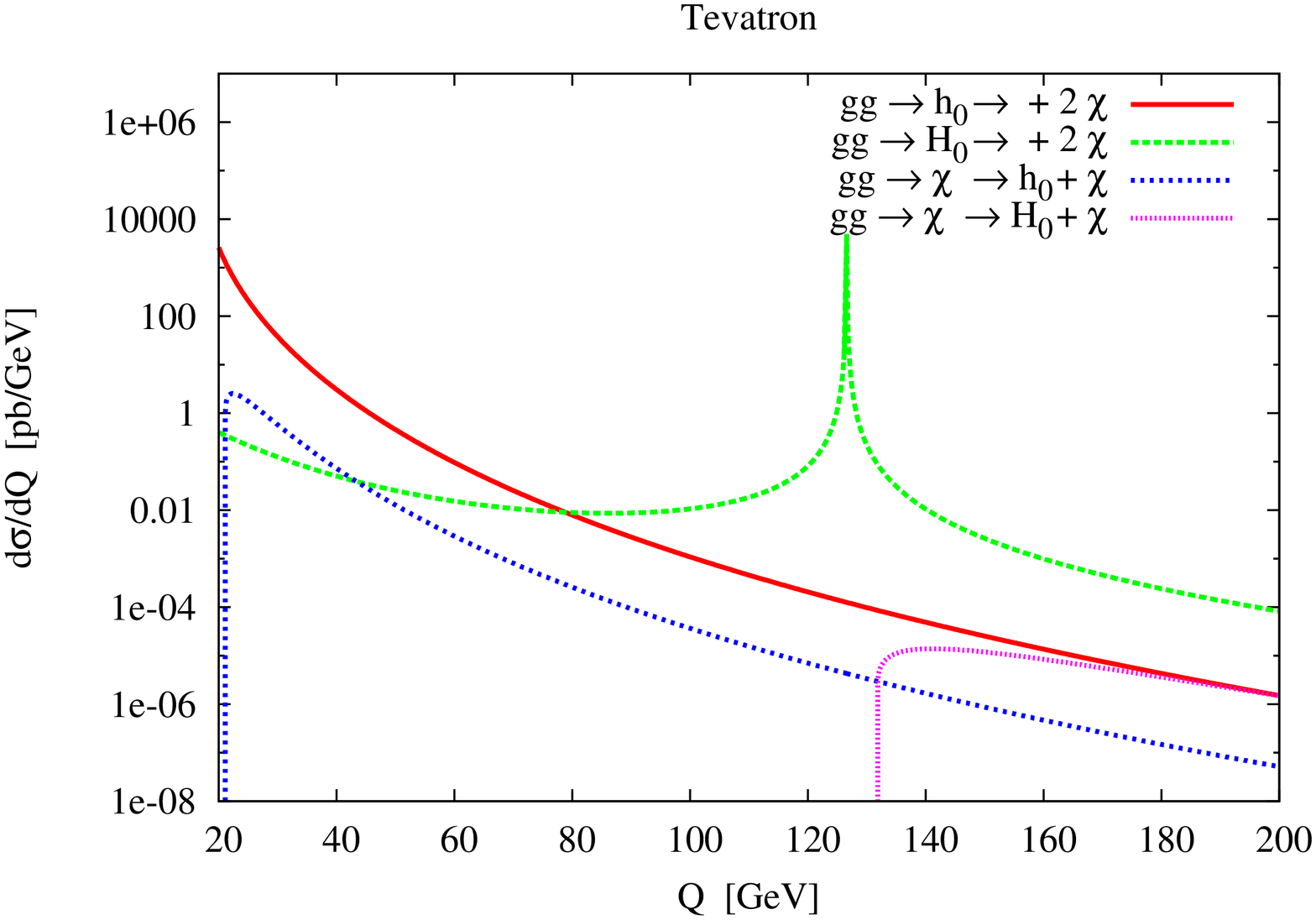}
\includegraphics[width=6cm, angle=-90]{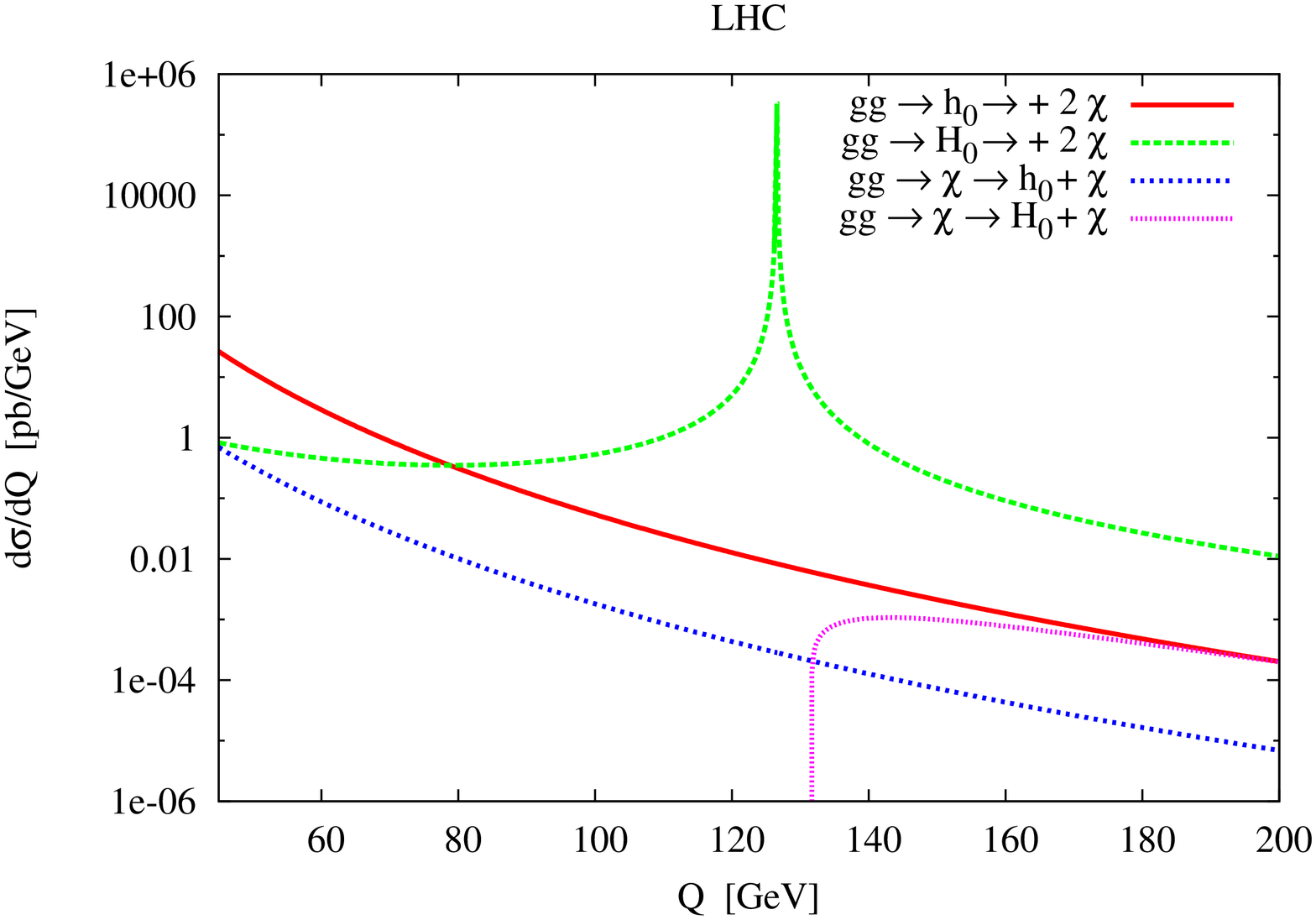}
\caption{\small $gg\rightarrow$ 2-scalar reactions mediated by trilinear vertices.}
\label{2scalars}
\end{figure}

Cascading channels can be easily obtained by combining trilinear 
($\chi\chi h_0$) and quadrilinear $(\chi^4)$ vertices, which are more involved. The results of this study are shown in Fig. \ref{2scalars} for the Tevatron and the LHC respectively.
 The plots presented in the two figures show sizeable rates which become large on the Higgs ($H_0$) resonance, chosen to be at 120 GeV. At the LHC the peak value of the cross section for $pp\to \chi\chi$, mediated by the $H_0$ is larger by a factor of about $10-100$ compared to the Tevatron and would be significant for future searches. 
 
\section{Conclusions} 
Axions derived from models containing anomalous extra $U(1)$ symmetries, if found at the LHC, would be of paramount 
theoretical importance. The search for this particle remains strongly tied to the possibility of uncovering an anomalous gauge boson at ATLAS and CMS in the next years and to the study of the trilinear gauge interactions \cite{Armillis:2007tb}. For an axion which is Higgs-like, the phenomenology is rather interesting and is characterized by sizeable production rates.

\begin{theacknowledgments}
This work is supported in part  by the European Union through the Marie 
Curie Research and Training Network Universenet (MRTN-CT-2006-035863). 
\end{theacknowledgments}

%\bibliography{TJJ_finalbib}

\end{document}